\documentclass[%
 aip,
 amsmath,amssymb,
 reprint,%
]{revtex4-1}

\usepackage{graphicx}% Include figure files
\usepackage{dcolumn}% Align table columns on decimal point
\usepackage{bm}% bold math

\usepackage{colordvi,color}
\usepackage[dvipsnames]{xcolor}  % More colors for text
\usepackage{float}
\usepackage[utf8]{inputenc}
\usepackage{soul,xcolor} 
\usepackage[version=4]{mhchem}

\usepackage{refstyle}
\let\eqref=\relax
\newref{eq}{name = Eq.~, names = Eqs.~, Name = Eq.~, Names = Eqs.~}
\newref{tab}{name = Tab.~, names = Tabs.~, Name = Tab.~, Names = Tabs.~}
\newref{fig}{name = Fig.~, names = Figs.~, Name = Fig.~, Names = Figs.~}
\newref{step}{name = step~, names = steps~, Name = Step~, Names = Steps~}
\newref{sec}{name = section~, names = sections~, Name = Section~, Names = Sections~}

\usepackage[unicode=true,pdfusetitle,
 bookmarks=true,bookmarksnumbered=false,bookmarksopen=false,
 breaklinks=false,pdfborder={0 0 1},backref=false,colorlinks=false]
 {hyperref}

\begin{document}
\title{Accurate Hellmann--Feynman forces from
  density functional calculations with augmented Gaussian basis sets}

\author{Shivesh Pathak}
\email{sapatha@sandia.gov}
\affiliation{Center for Computing Research, Sandia National Laboratories, United States of America}
\author{Ignacio Ema L\'{o}pez}%
\affiliation{Departamento de Química Física Aplicada, Universidad Aut\'{o}noma de Madrid, Spain}
\author{Alex J. Lee} 
\affiliation{Department of Chemical and Biological Engineering, University of New Mexico, United States of America}
\author{William P. Bricker} 
\affiliation{Department of Chemical and Biological Engineering, University of New Mexico, United States of America}
\author{Rafael L\'{o}pez Fern\'{a}ndez}%
\affiliation{Departamento de Química Física Aplicada, Universidad Aut\'{o}noma de Madrid, Spain}
\author{Susi Lehtola}%
\affiliation{Molecular Sciences Software Institute, Virginia Tech, United States of America}
\affiliation{Department of Chemistry, University of Helsinki, Finland}
\author{Joshua A. Rackers}
\affiliation{Center for Computing Research, Sandia National Laboratories, United States of America}

\newcommand{\la}{{\langle \, }}
\newcommand{\ra}{{\, \rangle}}
\newcommand{\bv}{{\, | \,}}
\newcommand{\alphavec}{\boldsymbol{\alpha}}

\newcommand{\sgbs}{$\sigma$ BS}
\newcommand{\sgdz}{$\sigma$DZ}
\newcommand{\sgnz}{$\sigma$NZ}
\newcommand{\sgtz}{$\sigma$TZ}
\newcommand{\sgdzhf}{$\sigma$DZHF}
\newcommand{\sgnzhf}{$\sigma$NZHF}
\newcommand{\sgtzhf}{$\sigma$TZHF}

\begin{abstract}
The Hellmann--Feynman (HF) theorem provides a way to compute forces
directly from the electron density, enabling efficient force
calculations for large systems through machine learning (ML) models
for the electron density. The main issue holding back the general
acceptance of the HF approach for atom-centered basis sets is the
well-known Pulay force which, if naively discarded, typically
constitutes an error upwards of 10 eV/\AA{} in forces. In this work, we
demonstrate that if a suitably augmented Gaussian basis set is used for density functional calculations, the Pulay force can be suppressed and HF
forces can be computed as accurately as analytical forces with
state-of-the-art basis sets, allowing geometry optimization and
molecular dynamics to be reliably performed with HF forces. Our
results pave a clear path forwards for the accurate and efficient
simulation of large systems using ML densities and the HF theorem.
\end{abstract}

\maketitle

\section{Introduction \label{sec:intro}}
A pressing issue in contemporary materials simulations is the accurate and efficient first principles calculation of atomic forces in large systems.
A representative class of such materials are protein molecules---some containing beyond 100,000 atoms---where accurate forces would allow for advancements in the understanding of processes such as protein folding.\cite{doi:10.1021/ar010018u, Trabanino2004-dr, PhysRevLett.90.238302, doi:10.1021/cr00023a010, PhysRevLett.89.108102, doi:10.1063/1.1329889, Dokholyan2001-oh}
Regarding periodic systems, materials with computational unit cells surpassing tens of thousands of atoms remain a challenge, while accurate simulation of the correlated electronic structure of van der Waals materials like twisted bilayer graphene would require force calculations on superlattices surpassing 40,000 atoms.\cite{PhysRevB.105.115141, PhysRevB.90.155451,  PhysRevB.93.235153, PhysRevResearch.2.043127}
Even with adaptations for increased efficiency,\cite{doi:10.1063/5.0004445, C5CP00437C, 10.3389/fchem.2020.589910} the most commonly used first principles method, density functional theory\cite{Hohenberg1964_PR_864, Kohn1965_PR_1133} (DFT), is unable to efficiently and accurately compute forces for such large-scale systems.

Machine learning (ML) has recently emerged as an effective solution to the seemingly intractable problem of accurate computation of properties of large systems.
The effectiveness of ML methods stems from their ability to extrapolate solutions from simple training data to more complex use cases: traditional first principles methods are only needed for generating the training data for the ML model, while the systems to which the ML techniques are eventually applied are typically orders of magnitude larger than the training configurations.
ML models are able to predict properties such as hopping parameters, \cite{PhysRevB.105.115141} potential energy surfaces, as well as forces \cite{doi:10.1021/acs.jpcc.6b10908, doi:10.1063/1.4966192, Handley2014} for large molecules and complicated solids.

While ML models are routinely trained to compute forces directly,\cite{doi:10.1021/acs.jpcc.6b10908} the Hellmann--Feynman (HF) theorem presents a promising alternative approach.
Under the Born--Oppenheimer (BO) approximation, the analytic expression for the force acting on nucleus $I$ situated at $\vec{R}_I$ is\cite{pulay1969ab}
\begin{equation}
    \vec{F}^E_{I} = - \frac {\partial E} {\partial \vec{R}_I} = -\frac{\partial \langle \Psi(\{\vec{R}\}) | \hat{H} | \Psi(\{\vec{R}\}) \rangle}{\partial \vec{R}_I} = \vec{F}_I^\text{HF} + \vec{F}_I^\text{Pulay}, \label{eq:analytic}
\end{equation}
where the HF term is
\begin{equation}
\vec{F}^\text{HF}_{I} = -\left\langle \Psi(\{\vec{R}\}) \left| \frac{\partial \hat{H}}{\partial \vec{R}_I} \right| \Psi(\{\vec{R}\}) \right\rangle \label{eq:hft}
\end{equation}
and the Pulay term\cite{pulay1969ab} originating from the geometry dependence of the basis set used to represent the electronic wave function $|\Psi\rangle$ is 
\begin{equation}
\vec{F}^\text{Pulay}_{I} = -2 \left\langle \Psi(\{\vec{R}\}) \Bigg| \hat{H} \Bigg| \frac{\partial \Psi(\{\vec{R}\})}{\partial \vec{R}_I} \right\rangle, \label{eq:pulay}
\end{equation}
where $\{ \vec{R} \}$ is the set of all nuclear positions and $\hat{H}$ is the electronic Hamiltonian.

It is easy to demonstrate that the HF force depends only on the electronic density $\rho$, as the only terms in the electronic Hamiltonian that depend on the nuclear coordinates $\{\vec{R}\}$ are nuclear-nuclear repulsion 
\begin{equation}
    \hat{H}^\text{nuc-nuc} = \sum_{IJ} \frac {Z_{I} Z_{J}} {|\vec{R}_{I}-\vec{R}_{J}|} \label{eq:nucnuc}
\end{equation}
and nuclear-electron attraction
\begin{equation}
    \hat{H}^\text{nuc-el} =- \sum_{I} Z_{I} \int \frac { \rho(\vec{r})} {|\vec{r}-\vec{R}_{I}|} {\rm d}^3 r, \label{eq:nucel}
\end{equation}
where $Z_I$ is the atomic number of nucleus $I$.
If one were able to appropriately suppress the Pulay force \eqref{pulay}, training an ML model for predicting accurate forces reduces to the simpler task of training an ML model for predicting accurate electronic densities, with forces computed using the predicted density via \eqref{hft}.
The computational expense of the latter task would be much lower than the prior since the training data would be much simpler to generate \cite{doi:10.1063/1.1286598, doi:10.1063/1.1562605}, yielding an appreciably efficient pipeline for computing accurate forces for large systems.

Perhaps the earliest musings on suppressing Pulay forces was by Pulay himself in ref.~\citenum{Pulay1977}, where he noted that the HF forces could be accurately computed using fixed basis functions if the basis could accurately describe the derivatives of the wave functions $|\partial \Psi /\partial \vec{R}_I\rangle$.
A few years later, Nakatsuji and coworkers \cite{NAKATSUJI1980340, Nakatsuji1981, doi:10.1063/1.444234} began constructing such basis sets, where derivatives of the basis functions were included in the orbital basis.
The resulting "family" style basis sets were found to yield reasonably accurate HF forces, equilibrium geometries, and force constants for small molecules.

However, Pulay levied criticism on this approach\cite{doi:10.1063/1.446089} for (i) the computational expense of including large numbers of core polarization functions that are necessary for accurate HF forces\cite{Pulay1977} as well as (ii) the poor accuracy of the approach of Nakatsuji and coworkers for molecular geometries compared to standard basis sets with the analytical derivative approach.
Nakatsuji \textit{et al.} rebutted the criticism  by stating that even though the computational costs are higher, the results are easier to interpret chemically thanks to the fulfillment of the HF theorem.\cite{doi:10.1063/1.446090}

Pulay's views have been since widely adopted by the quantum chemistry community: HF forces have practically vanished from consideration.
However, it is important to note that the pioneering studies by Pulay and Nakatsuji and coworkers were limited to small basis sets due to the meager computational resources available at the dawn of quantum chemistry.
Small basis sets are known to be unreliable for chemical applications.
Even if the Pulay term of \eqref{pulay} is suppressed in such a basis, the electron density $\rho$ may still have large errors, which lead to large errors in the HF force, as well.
Modern applications of quantum chemistry routinely use extended basis sets of polarized triple-$\zeta$ quality or higher, which were not tractable at the time of the early studies of Pulay, Nakatsuji and coworkers.
The applicability of the old ideas of Pulay, Nakatsuji and coworkers thus deserve revisiting.

Much more recently, \citet{rico2007generation} proposed a new "family" style basis set for Slater-type orbitals with preliminary tests demonstrating noticeably attenuated Pulay forces and accurate electronic structures without the excessive overheads suggested by Pulay.
Unresolved by \citeauthor{rico2007generation}, however, are the questions of applicability of these forces in relevant tasks like geometry optimization and molecular dynamics.
In this work, we follow \citeauthor{rico2007generation} and demonstrate that specially optimized atom-centered Gaussian basis sets yield HF forces with an accuracy comparable to analytic forces computed with state-of-the-art Gaussian basis sets within DFT, and that they can be used to get state-of-the-art accuracy in applications of geometry optimization and molecular dynamics.
We believe our results provide a promising avenue for the use of HF forces in the accurate and efficient ML modelling of forces for large systems.

The organization of this work is the following.
In \secref{hfbasis}, we describe the approach that is used to augment the $\sigma$NZ basis sets (N = \textbf{S}ingle, \textbf{D}ouble, \textbf{T}riple) of \citet{Ema2022} to reduce the Pulay term in \eqref{analytic}, yielding the $\sigma$NZHF basis sets whose accuracy is demonstrated in this work.
\Secref{compute} includes computational details for DFT, as well as geometry optimization and molecular dynamics (MD) methods employed in the rest of the manuscript, followed by the results of these computations in \secref{results}.
We finish in \secref{conclusions} with a short summary and conclusions, highlighting the efficient pipeline for building ML force models enabled by our work.

The organization of the results, \secref{results}, is as follows.  We begin the  demonstration of the accuracy of HF forces by computing force curves for the CO molecule in \secref{diatomic}.
In \secref{benchmark}, we carry out an in-depth analysis of total energies, analytical forces, and HF forces over a standard benchmark set of molecules,\cite{https://doi.org/10.1002/jcc.540140910} and follow-up with a similar analysis for DNA fragments in \secref{dnafragments}.
Our results show that the HF forces computed using the $\sigma$DZHF and $\sigma$TZHF basis sets yield similar accuracy to analytical forces computed using the size-matched $\sigma$NZ \cite{Ema2022} and \mbox{(aug-)pcseg-N} \cite{doi:10.1021/ct401026a} basis sets.
With this in mind, we follow up in \secref{optimization} with a study of geometry optimization on the benchmark molecule set of ref.~\citenum{https://doi.org/10.1002/jcc.540140910} in \secref{benchmark}, finding that optimizing geometries with HF forces computed using the $\sigma$DZHF basis yields less than 1\% error compared to geometries optimized with analytic forces and the pcseg-2 basis.
As a final application, we conduct a 100 fs room-temperature Born--Oppenheimer molecular dynamics (BOMD) calculation on a single ethanol molecule using HF forces and the $\sigma$DZHF basis in \secref{md}, finding excellent agreement in both total energies and configurations when compared to a BOMD calculation run with analytic forces computed using the pcseg-2 basis.

\section{Basis set augmentation \label{sec:hfbasis}}

% \subsection{Sizes and composition of $\sigma$ basis sets}
The $\sigma$NZHF basis sets used in this work are generated by augmentation of the family-style $\sigma$NZ basis sets (\textbf{N} = \textbf{S}ingle, \textbf{D}ouble, \textbf{T}riple) of \citet{Ema2022}.
Below we provide a sufficient overview of the basis set construction procedure for the reader, with additional implementation details in the Supplementary Material.

We use the strategy of \citet{rico2007generation}  to iteratively augment the $\sigma$NZ basis sets.
Functions are added into the basis set, until the error $\Delta_\lambda$ in the basis set projection of the nuclear forces of the basis functions (see Eq.~6 of ref.~\citenum{rico2007generation}) becomes suitably small, as determined by a parameter $\epsilon$, $\Delta_\lambda < \epsilon$.
As a result of this procedure, the Pulay forces are reduced to the order of $\epsilon$.

The $\sigma$NZ basis sets were chosen for this work, since the $\sigma$NZ basis sets consist of contractions of primitive Gaussian-type orbitals (pGTOs) whose exponents are shared by several angular momenta: if a given primitive appears multiplied by a spherical harmonic of quantum number $l = L$, analogous functions with the same exponent are also present for $0 \leq l < L$.
The use of shared exponents results in the inclusion of a great deal of the derivative space in the basis, simplifying the efforts of this work.
However, the general methodology used in this work based on that of ref.~\citenum{rico2007generation} could also be used to augment other types of basis sets for fulfillment of the HF theorem.
We note that the $\sigma$NZ basis sets of \citet{Ema2022} have been optimized following the general lines of the procedure of \citet{doi:10.1063/1.456153} by minimizing the configuration interaction singles and doubles (CISD) energy of gas-phase atoms; the composition of the $\sigma$NZ basis sets is shown in \tabref{tabl}.

\begin{table}
\caption{Number of primitive and contracted basis functions for the $\sigma$NZ \cite{Ema2022} and $\sigma$NZHF (this work) basis sets for the considered first row and second row atoms.
$N_\text{exp}$ denotes the number of unique exponents in the basis set.
}
\resizebox{\columnwidth}{!}{%
\begin{tabular}{|l|c|rl|rl|}
\hline 
Basis  & $N_\text{exp}$
& $\#$ & Primitives & $\#$ & Contracted \\
\hline
\hline
 & \multicolumn{5}{c|}{H atom} \\
\hline
$\sigma$SZ     & 10 & 10 & (10s)                      & 1 & [1s]             \\
$\sigma$DZ     & 10 & 19 & (10s,  3p)                 & 5 & [2s,  1p]        \\
$\sigma$TZ     & 10 & 37 & (10s,  4p, 3d)             & 14 & [3s,  2p, 1d]     \\
\hline
$\sigma$SZHF   & 10 & 40 & (10s,  10p)                & 4 & [1s,  1p]          \\
$\sigma$DZHF   & 10 & 55 & (10s,  10p, 3d)            & 17 & [3s,  3p, 1d]      \\
$\sigma$TZHF   & 12 & 85 & (11s,  11p, 4d, 3f)        & 46 & [6s,  6p, 3d, 1f]   \\
\hline
 & \multicolumn{5}{c|}{C, N, O and F atoms} \\
\hline
$\sigma$SZ     & 15 & 45 & (15s, 10p)                & 5 & [2s,  1p]             \\
$\sigma$DZ     & 15 & 60 & (15s, 10p, 3d)            & 14 & [3s,  2p, 1d]         \\
$\sigma$TZ     & 15 & 86 & (15s, 10p, 4d, 3f)        & 30 & [4s,  3p, 2d, 1f]     \\
\hline
$\sigma$SZHF   & 15 & 110 & (15s, 15p, 10d)           & 17 & [3s,  3p, 1d]         \\
$\sigma$DZHF   & 15 & 131 & (15s, 15p, 10d, 3f)       & 46 & [6s,  6p, 3d, 1f]     \\
$\sigma$TZHF   & 17 & 169 & (16s, 16p, 10d, 4f, 3g)   & 87 & [8s,  8p, 5d, 3f, 1g]        \\
\hline
 & \multicolumn{5}{c|}{P, S and Cl atoms} \\
\hline
$\sigma$SZ     & 19 & 79 &(19s, 15p)                 &  9 & [3s,  2p]            \\
$\sigma$DZ     & 19 & 79 &(19s, 15p, 3d)             & 18 & [4s,  3p,  1d]            \\
$\sigma$TZ     & 19 & 95 & (19s, 15p, 4d, 3f)        & 34 & [5s,  4p,  2d, 1f]        \\
\hline
$\sigma$SZHF   & 19 & 151 & (19s, 19p, 15d)           & 30 & [5s,  5p,  2d]        \\
$\sigma$DZHF   & 19 & 172 & (19s, 19p, 15d, 3f)       & 59 & [8s,  8p,  4d, 1f]        \\
$\sigma$TZHF   & 21 & 210 & (20s, 20p, 15d, 4f, 3g)   & 100 & [10s, 10p, 6d, 3f, 1g]    \\
\hline
\end{tabular}
}
\label{tab:tabl}
\end{table}

To construct a basis with a high degree of fulfillment of the HF theorem, we extend the $\sigma$NZ basis sets with functions corresponding to the occupied basis functions' derivatives. 
As shown in Eq.~16 of the Supplementary Material, the derivative of a pGTO with a given $l > 0$ yields three functions with the same exponent as in the pGTO: one function corresponding to $l+1$, another to $l-1$, and one function to $l-1$ but bearing an additional factor $r^2$.
As the present basis sets employ spherical functions only, the radial factors included in the pGTOs for angular momentum $l$ is $r^l$.
Because of this, we remove the additional functions with $r^2$ from consideration, yielding what we call a {\it reduced set of primitives} that span the {\it reduced space}.

Our goal is to iteratively improve the \textit{reduced space} so that the distance between it and the \textit{reference space}---the space spanned by contractions of $\sigma$NZ and their derivatives---becomes smaller than the used threshold $\epsilon = 10^{-3}$.
The procedure proceeds as follows.
\begin{enumerate}
    \item[0.] Choose the initial basis set, yielding a \textit{reduced space} and a \textit{reference space}.
    \item Compute $\Delta_\lambda$ with the given reduced space and reference space. \label{step:computedelta}
    \item If $\Delta_\lambda < \epsilon$, stop. \label{step:stop}
    \item Otherwise, add additional primitive functions to the reduced set and go back to \stepref{computedelta}.
\end{enumerate}
Details on computing $\Delta_\lambda$ for Gaussian basis functions as well as the decision criteria for exponents and angular momenta of the added pGTOs can be found in Sections~2--5 of the Supplementary Material.

Once the improved {\it reduced set of primitives} has been formed, we carry out an expansion of the functions of the {\it reference set} in an orthogonal basis of the {\it reduced set of primitives} to form the final contracted $\sigma$NZHF basis sets.
We find that projecting into the full {\it reduced set of primitives} is generally unnecessary.
Rather, by choosing a suitable subspace of the {\it reduced space} to project into, we can reduce the number of final contracted basis functions in $\sigma$NZHF up to 20\% for heavier atoms.
Details of the iterative approach for finding a suitable subspace can be found in Section~5 of the Supplementary Material.

The $\sigma$NZHF basis sets resulting from this procedure are the final HF basis sets used in this work.
The compositions of the $\sigma$NZHF basis sets are shown in \tabref{tabl} in terms of primitive and contracted functions; the compositions of the original $\sigma$NZ basis sets are also included for comparison.
It should be noted that additional pGTOs (namely, $s$ and $p$ primitives) were only required for the $\sigma$TZHF basis sets.
The $\sigma$NZHF basis sets for H, C, N, O, F, P, S and Cl are provided in the Supplementary Material and can be found on the Basis Set Exchange \cite{doi:10.1021/acs.jcim.9b00725, doi:10.1021/ci600510j}.

\section{Computational details \label{sec:compute}}
We compute forces from first principles to conduct geometry optimization and carry out BOMD simulations for a wide range of molecular systems.
The electronic first principles problem is solved using DFT with the PBE0 functional \cite{doi:10.1063/1.478401, doi:10.1063/1.478522} from Libxc~\cite{LEHTOLA20181}.
DFT calculations in \secref{diatomic, benchmark, dnafragments} were carried out with a custom version of the Psi4 package~\cite{doi:10.1063/5.0006002} available at \url{https://github.com/JoshRackers/psi4}, while all other DFT calculations were done using the PySCF package \cite{https://doi.org/10.1002/wcms.1340, doi:10.1063/5.0006074}.
Density fitting was employed with the def2-universal-jkfit auxiliary basis of \citet{https://doi.org/10.1002/jcc.20702}. 
A (75, 302) quadrature grid for the DFT functional was used for all atoms in Psi4, and all atoms except hydrogen in PySCF, for which a (50, 302) quadrature grid was used instead.

Geometry optimization was carried out using the geomeTRIC\cite{doi:10.1063/1.4952956} solver within PySCF.
We use the default geomeTRIC convergence criteria for optimization, namely a energy convergence criterion of $10^{-6} E_h$, force maximum and RMS values of $3 \times 10^{-4} E_h/a_0$  and $4.5 \times 10^{-4} E_h/a_0$, respectively,  with $a_0$ the Bohr radius and configuration deviation maximum and RMS values of $1.2 \times 10^{-3} $ \AA{} and $1.8 \times 10^{-3}$  \AA{}, respectively.
BOMD calculations are carried out in the NVE ensemble, also within the PySCF package.
The Verlet integrator \cite{PhysRev.159.98} is used out to a final time of 100 fs using 1 fs time steps, and an initial randomized velocity is given to the system to maintain a time-averaged temperature $\sim$ 300K.
For clarity, we note that when HF forces are employed for optimization, convergence criteria such as the force maximum and RMS force refer to the computed HF forces; when using HF forces for MD, all accelerations are computed using HF forces only.

\section{Results \label{sec:results}}
\subsection{Diatomic force curve for CO \label{sec:diatomic}}
\begin{figure*}
    \centering
    \includegraphics[width=\textwidth]{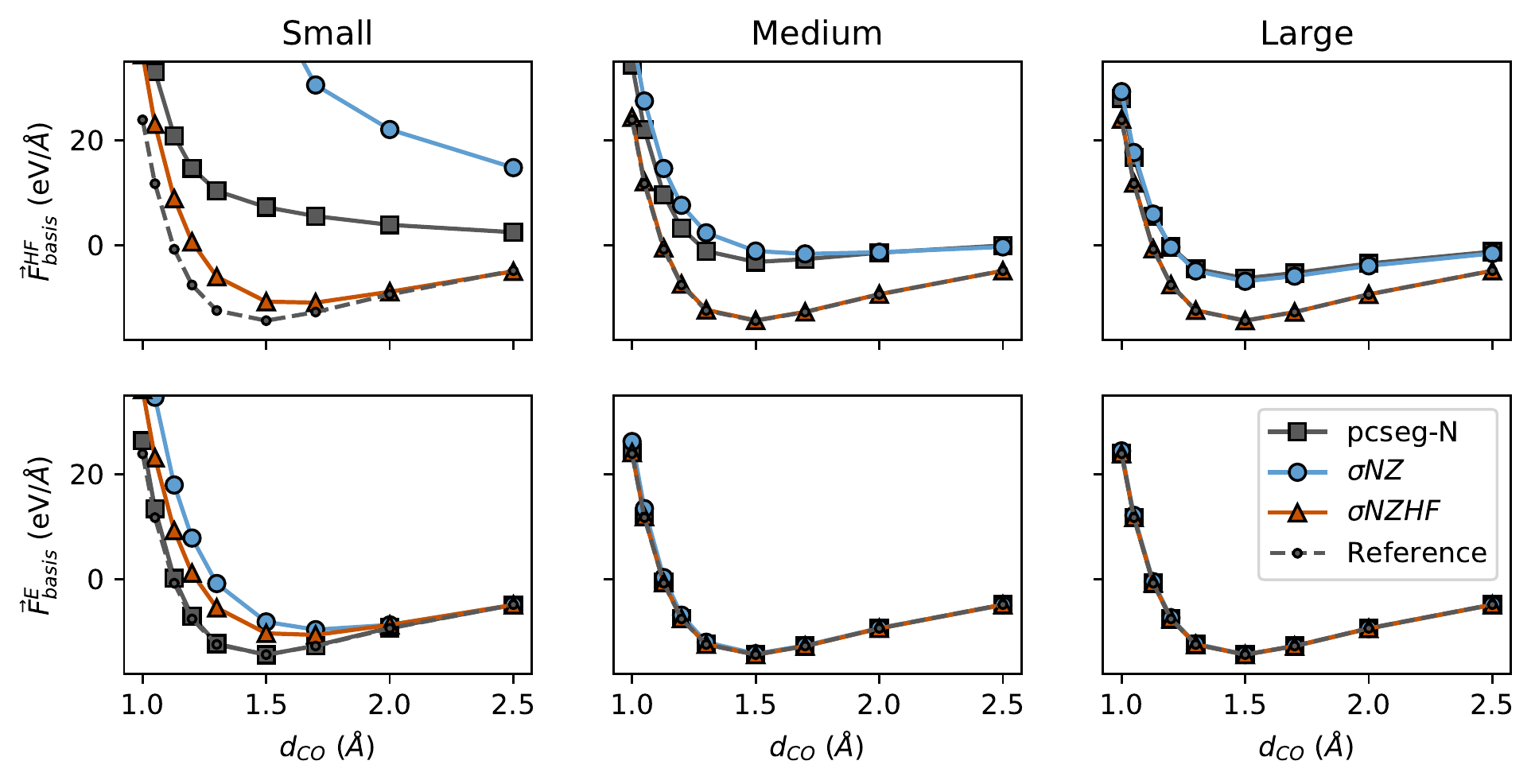}
    \caption{Diatomic force curves for CO molecule computed as a function of bond length $d_{CO}$.
    The force shown is the force along the bond, namely $\vec{F}_\text{CO} = (\vec{F}_\text{C} - \vec{F}_\text{O})/2$.
    Columns are delineated by the basis set size classified within \tabref{tabl2}, and rows by the force computed: analytic or HF.
    The dashed Reference curve is a reference calculation for the reader, computed using the large pcseg-4 basis and analytic forces.}
    \label{fig:curves}
\end{figure*}

We begin by computing a force curve for the CO molecule as a simple demonstration of the accuracy of HF forces in DFT using the $\sigma$NZHF basis, shown in \figref{curves}; this molecule was also examined by \citet{doi:10.1063/1.444234} with Hartree--Fock and small Pople-style basis sets.
To understand the effect of the added basis functions to fulfill the HF theorem, we also compute analytic and HF forces using the $\sigma$NZ basis sets that were the starting point for the $\sigma$NZHF basis sets.
Furthermore, to make a fair comparison to the state-of-the-art in basis set development, we computed energies and forces using size-matched pcseg-N basis sets \cite{doi:10.1021/ct401026a} that are optimized for accurate DFT total energies.
A reference curve computed using the analytic force and the pcseg-4 basis set is also shown in each panel. 
As a point of reference, \tabref{tabl2} presents the comparison of basis set sizes for the $\sigma$NZHF, $\sigma$NZ and pcseg-N basis sets for a single CO molecule.

We begin by comparing the HF force computed using the $\sigma$NZ, $\sigma$NZHF, and pcseg-N basis sets over the benchmark molecule set, shown in the first row of \figref{curves}.
Large deviations from the reference forces can be observed in the results obtained with the $\sigma$NZ and pcseg-N basis sets.
These large differences arise from the unsuitability of standard basis sets for computing HF forces: even the quadruple-$\zeta$ pcseg-3 basis set (shown in the rightmost column) exhibits considerable errors in the HF force.
In contrast, we find that the $\sigma$NZHF basis sets of this work yield force curves which agree well with the reference calculation.
While the curve computed with the $\sigma$SZHF basis set differs visually from the reference curve for reasons discussed in the following paragraph, the $\sigma$DZHF and $\sigma$TZHF curves are in excellent agreement with the reference data.
These results demonstrate the efficacy of the optimized $\sigma$NZHF sets of this work in reducing the Pulay forces.

Analytic forces are compared in the second row of \figref{curves}.
The larger $\sigma$DZHF and $\sigma$TZHF basis sets again yield good agreement with the reference calculation, and afford similar accuracy to the size-matched pcseg-N basis sets that have been designed for DFT calculations.
The smallest basis set $\sigma$SZHF, however, yields larger errors than pcseg-1 when computing analytic derivatives.

The comparison of the HF and analytic forces for $\sigma$SZHF, shown in the upper and lower panels of the left-most column of \figref{curves}, respectively, shows that the HF force is an accurate approximation of the analytic gradient in the $\sigma$SZHF basis set, that is, the Pulay forces have been successfully made negligible by the design of the basis set.
Combined with the inaccuracy of the analytic gradient in the $\sigma$SZHF calculation, this indicates that the $\sigma$SZHF basis set is simply too small for quantitative electronic structure calculations, like its parent, the minimal $\sigma$SZ basis set.
The error in the HF force arises mostly from the limitations in the description of the electron density in the small basis set, as discussed in the Introduction.

Although data for $\sigma$SZHF basis sets is included throughout this work as a point of reference, $\sigma$SZHF results will not be discussed in the remainder of this work, as the same remarks on the inherently limited accuracy of small basis sets apply throughout the discussion.
The larger $\sigma$DZHF and $\sigma$TZHF basis sets, however, not only provide accurate analytical gradients, but are also useful for HF force calculations.
These results illuminate the pivotal difference between the pioneering attempts at HF forces in the literature\cite{NAKATSUJI1980340, Nakatsuji1981, doi:10.1063/1.444234, doi:10.1063/1.446089, doi:10.1063/1.446090} and the approaches pursued in this work: although small basis sets only afford results of limited quality, larger basis sets allow the reproduction of accurate HF forces while still being compact enough to enable routine calculations on existing computing platforms.

As a last point we discuss the total force acting on the system, $\vec{F}_\text{tot} = (\vec{F}_C + \vec{F}_O)$, which should be zero for a calculation with no external forces.
The analytic DFT force calculations will not have exactly vanishing total force, as the finite integration grid for the DFT functional breaks continuous spatial translation symmetry and thereby total momentum conservation is lost.
HF forces computed within DFT will suffer from the same issue, compounded with the fact that discarding the Pulay force of \eqref{pulay}, no matter how small, will lead to an additional systematic error in the total force.
For analytic forces computed with the pcseg-3 basis we find $|\vec{F}_\text{tot}| \sim 10^{-5}$ eV/\AA{}, while for HF forces computed with the $\sigma$TZHF basis, we find $|\vec{F}_\text{tot}| \sim 10^{-3}$ eV/\AA{}.
We believe that both of these total force magnitudes are sufficiently small for practical and accurate force calculations.
We would also like to note that errors in the HF forces introduced by discarding the suppressed Pulay force cannot break discrete symmetries of molecular systems, unless the electron density used to compute the HF force also breaks the symmetry.
This ensures, for example, that out-of-plane forces for planar molecules always vanish, unless the electron density breaks the planar symmetry by an asymmetricity around the plane.

\subsection{Force comparison at fixed geometries of a benchmark set \label{sec:benchmark}}
We follow up our force curve demonstration with a more robust analysis of force errors over a benchmark set of small- to medium-size organic molecules taken from \citeauthor{https://doi.org/10.1002/jcc.540140910}'s seminal work on geometry optimization.\cite{https://doi.org/10.1002/jcc.540140910}
In this study we consider eight molecules from the benchmark set: water, ammonia, ethane, acetylene, allene, methylamine, hydroxysulphane, and ethanol.
For each molecule, the non-equilibrium configurations stated as "starting configurations" in \citeauthor{https://doi.org/10.1002/jcc.540140910}'s work are used to compute HF forces, analytic forces and total energies.

%the Treutler \cite{doi:10.1063/1.469408} scheme

\begin{table}
\resizebox{0.8\columnwidth}{!}{%
\begin{tabular}{|l | c | c | c | c | c | c |}
\hline
Category & Basis & Size & Basis & Size & Basis & Size \\
\hline
\hline
Small & $\sigma$SZ & 10 & $\sigma$SZHF & 34 & pcseg-1 & 28 \\
Medium & $\sigma$DZ & 28 & $\sigma$DZHF & 92 & pcseg-2 & 60 \\
Large & $\sigma$TZ & 60 & $\sigma$TZHF & 174 & pcseg-3 & 120\\
\hline
Reference & & & & & pcseg-4 & 202 \\
\hline
\end{tabular}
}
\caption{Comparison of basis set sizes for a single CO molecule. The $\sigma$NZHF series of this work are contrasted to the $\sigma$NZ and pcseg-N series used in this work.
The "Small," "Medium," and "Large" categories are used in this work for comparison between various basis sets of similar size.}
\label{tab:tabl2}
\end{table}

\begin{figure*}
\includegraphics[width = \textwidth]{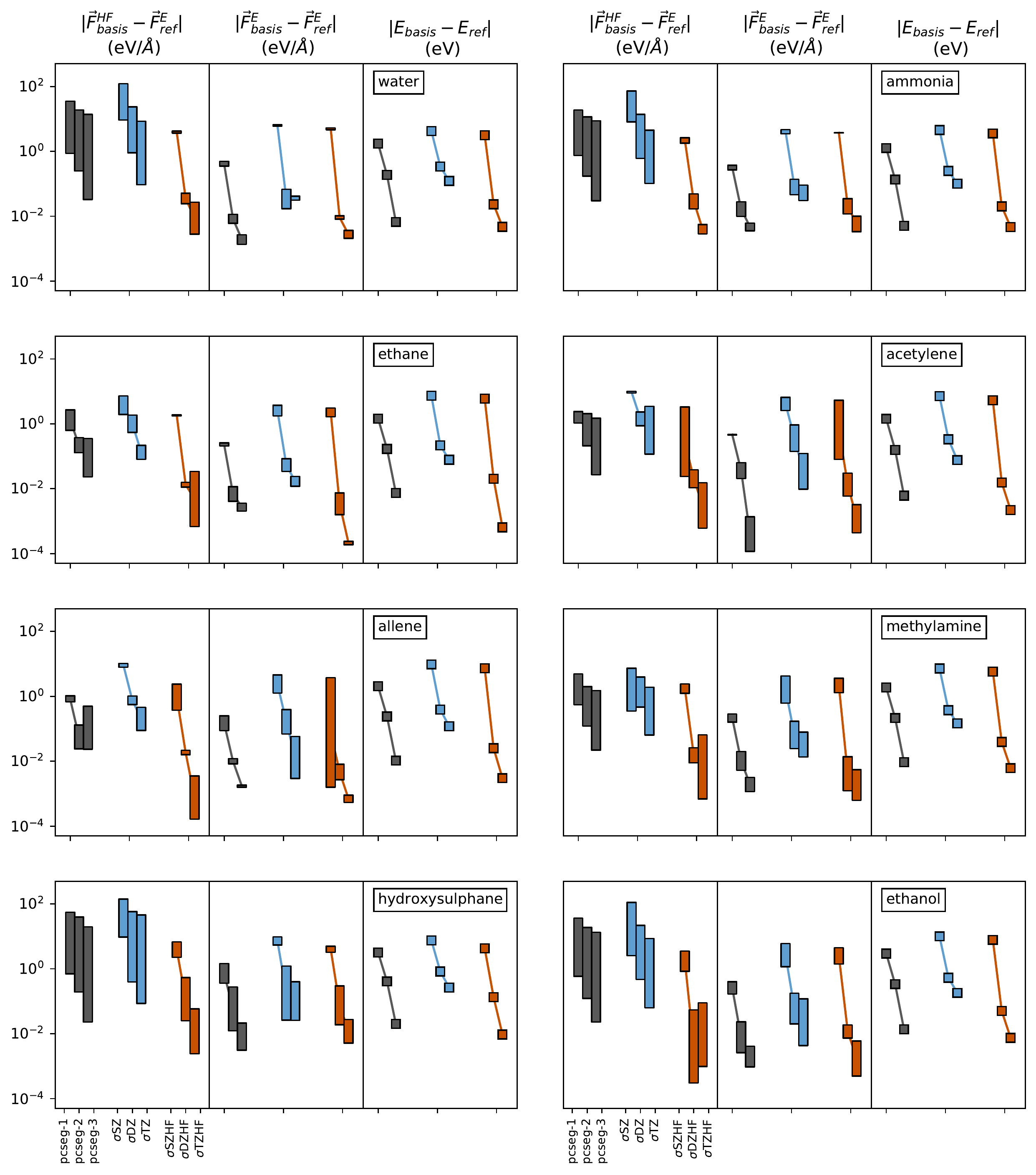}
\caption{Comparison of HF forces, analytic forces, and total energies for various small- to medium-size benchmark molecules.
Each set of three panels are data for a particular molecule, with one panel for each physical quantity computed. 
The x-axis denotes the basis set used in the calculation, namely the three basis series with Small, Medium and Large sets defined in \tabref{tabl2}.
Since force vectors are computed on each atomic center of a given molecule, this results in multiple data points per force calculation; these points are represented by bars denoting the range of the force errors.
In contrast, there is only one total energy per molecule.
Absolute errors are plotted relative to analytic forces and energies computed using the pcseg-4 basis.
}
\label{fig:force}
\end{figure*}

We begin by comparing the HF force computed using the $\sigma$NZ, $\sigma$NZHF, and pcseg-N basis sets over the benchmark set.
The errors in the HF forces ($\vec{F}_\text{basis}^{\text{HF}}$) compared to analytic forces computed in the pcseg-4 basis ($\vec{F}_\text{ref}^{\text{E}}$) are shown in the first panel of each sub-figure of \figref{force}.
Since one force vector is obtained for each atomic center in the molecule, the distribution of force errors is visualized by bars that covers the range of errors from maximum to minimum over all the atoms in the molecule.

We find that the $\sigma$DZHF and $\sigma$TZHF basis sets yield HF forces with errors that are nearly two orders of magnitude smaller than those of the similarly sized $\sigma$NZ and pcseg-N basis sets, once again demonstrating the ability of the $\sigma$NZHF basis sets to reduce Pulay forces.
The poorer performance of $\sigma$SZHF can again be attributed to its insufficient size, as discussed in \secref{diatomic}, and its results will not be discussed in detail.

Analytic forces are compared in the second panel of each sub-figure in \figref{force}.
The larger $\sigma$DZHF and $\sigma$TZHF basis sets once again afford a 5-fold to 10-fold reduction in the force error relative to the $\sigma$NZ, and similar accuracy to the pcseg-N basis sets which have been designed for DFT calculations.
A further important point is the ability of the $\sigma$NZHF basis sets in computing the total energy; these results are presented in the third panel of the sub-figures of \figref{force}.
The $\sigma$NZHF basis sets yield fast convergence to the basis set limit with an accuracy that is surprisingly similar to or even better than that of the pcseg-N basis sets that have been optimized for DFT calculations.

The two largest $\sigma$NZHF (and pcseg-N) basis sets both afford errors 1--2 orders of magnitudes smaller than the similarly sized $\sigma$NZ basis sets.
This is somewhat surprising, given that the $\sigma$NZHF basis sets are generated from the $\sigma$NZ basis sets by inclusion of basis function derivatives. 
The inclusion of such derivatives leads to considerable improvements in the accuracy of DFT total energies, in addition to better satisfaction of the HF theorem.
The HF basis sets may thus be useful for accurate computations of total energies and analytical derivatives, as well.

\subsection{Force comparison at fixed geometries on DNA fragments
\label{sec:dnafragments}}

\begin{figure*}
\includegraphics[width=\textwidth]{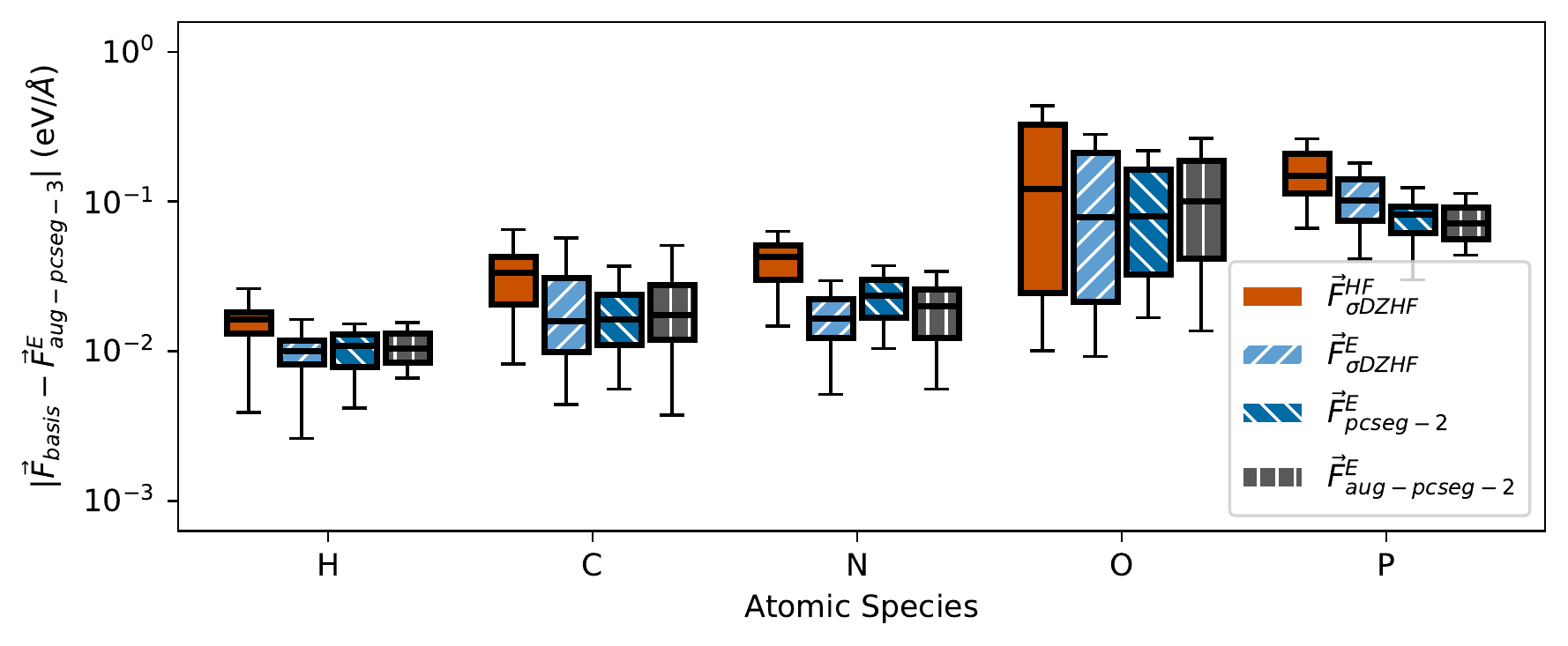}
\caption{Error in the computed forces for the DNA structures, with the $\sigma$DZHF, pcseg-2 and aug-pcseg-2 basis sets.
The reference force was computed using the aug-pcseg-3 basis with the analytic force technique with RMS magnitudes 0.82 eV/\AA{}, 5.33 eV/\AA{}, 5.56 eV/\AA{}, 4.41 eV/\AA{} and 16.26 eV/\AA{} for the H, C, N, O and P atoms respectively.
}
\label{fig:dna}
\end{figure*}

As a final demonstration of the accuracy of HF forces using the $\sigma$NZHF basis set, we computed HF and analytical forces on small DNA fragments.
DNA fragments are formed of more elements than present in the benchmark set of \secref{benchmark}, containing P atoms in addition to H, C, N and O. 
Therefore, the DNA calculations allow us to determine the accuracy of HF forces computed with the $\sigma$NZHF basis sets in a system containing second row atoms. 

DNA fragment geometries were obtained from MD simulations performed in a previous study~\cite{lee_rackers_bricker_2022} using the Amber 20 package~\cite{AMBER} and the BSC1 force field~\cite{Ivani2016}. 
DNA was modeled as 12 base pair strands (``12-mers") in the canonical B-form \cite{Richmond2003}, the most common structural form of DNA. 
To ensure adequate sampling of DNA's conformational space, MD simulations were run on four different 12-mer base sequences. 
Solvent molecules were modeled explicitly as TIP3P water~\cite{Jorgensen1983} with a \ce{Mg^{2+}} and \ce{Cl-} counterion concentration of about 100 mmol/L. 

Initial DNA structures were first minimized and then allowed to heat up from 0 K to 300 K for 40 ps. 
After heating, 50 ns production runs were performed in the NPT ensemble at 1 atm and 300 K. We used the Langevin thermostat with a collision frequency of 1 ps$^{-1}$, the Berendsen barostat with a relaxation time of 2 ps, and a time step of 2 fs.
For each DNA structure, three simulations were performed with different starting trajectories for a combined simulation time of 150 ns. Fragment geometries were constructed from production run snapshots by extracting the central two base pairs of the 12-mer sequence and stripping away the rest of the molecule.  

From the MD snapshots of DNA configurations, we analyzed ten configurations of six types of DNA fragments for a total of 60 structures. The six fragments include the four nucleotides (base + sugar + phosphate) with each of the bases (A, C, G, and T), and the two base pair structures (A-T and C-G), each fragment containing around 35 atoms. 
Note that the nucleotides are negatively charged due to the phosphate group whereas the base pair structures are neutral.

\begin{table*}
    \centering
    \resizebox{1.2\columnwidth}{!}{
    {\renewcommand{\arraystretch}{1.4}
    \begin{tabular}{|c|c|c|c|c|}
    \hline
     System   & Optim. & $|d - d_{\text{ref}}|$ (m\AA) & $|\theta - \theta_{\text{ref}}|$ ($^\circ$) & $|\phi  - \phi_{\text{ref}} |$ ($^\circ$) \\
     \, & \, & Min --- Max & Min --- Max & Min --- Max \\
     \hline
     \hline 
    water & $\vec{F}^{E}_{\text{pcseg-2}}$ & 0.017 --- 0.017 & 0.045 --- 0.045 &  \\
    & $\vec{F}^\text{HF}_{\sigma \text{DZHF}}$ & 0.579 --- 0.579 & 0.123 --- 0.123 &  \\
    \hline 
    ammonia & $\vec{F}^{E}_{\text{pcseg-2}}$ & 0.247 --- 0.247 & 0.230 --- 0.230 &  \\
    & $\vec{F}^\text{HF}_{\sigma \text{DZHF}}$ & 0.707 --- 0.707 & 0.228 --- 0.228 &  \\
    \hline 
    ethane & $\vec{F}^{E}_{\text{pcseg-2}}$ & 0.369 --- 0.504 & 0.004 --- 0.004 & 0.0001 --- 0.0001 \\
    & $\vec{F}^\text{HF}_{\sigma \text{DZHF}}$ & 0.462 --- 1.334 & 0.009 --- 0.010 & 0.0001 --- 0.0001 \\
    \hline 
    acetylene & $\vec{F}^{E}_{\text{pcseg-2}}$ & 0.216 --- 0.696 &  &  \\
    & $\vec{F}^\text{HF}_{\sigma \text{DZHF}}$ & 0.207 --- 0.695 &  &  \\
    \hline 
    allene & $\vec{F}^{E}_{\text{pcseg-2}}$ & 0.238 --- 0.410 & 0.001 --- 0.037 & 0.0001 --- 0.0001 \\
    & $\vec{F}^\text{HF}_{\sigma \text{DZHF}}$ & 0.451 --- 0.727 & 0.001 --- 0.003 & 0.0001 --- 0.0002 \\
    \hline 
    methylamine & $\vec{F}^{E}_{\text{pcseg-2}}$ & 0.126 --- 0.443 & 0.016 --- 0.066 & 0.001 --- 0.049 \\
    & $\vec{F}^\text{HF}_{\sigma \text{DZHF}}$ & 0.306 --- 1.985 & 0.023 --- 0.084 & 0.001 --- 0.025 \\
    \hline 
    hydroxysulphane & $\vec{F}^{E}_{\text{pcseg-2}}$ & 0.021 --- 5.756 & 0.226 --- 0.326 & 0.490 --- 0.490 \\
    & $\vec{F}^\text{HF}_{\sigma \text{DZHF}}$ & 0.516 -- 11.217 & 0.196 --- 0.600 & 0.676 --- 0.676 \\
    \hline 
    ethanol & $\vec{F}^{E}_{\text{pcseg-2}}$ & 0.093 --- 0.415 & 0.002 --- 0.087 & 0.0001 --- 0.049 \\
    & $\vec{F}^\text{HF}_{\sigma \text{DZHF}}$ & 0.137 --- 3.606 & 0.001 --- 0.216 & 0.0001 --- 0.118 \\
    \hline 
    \end{tabular}
    }
    }
    \caption{Errors for optimized geometries using HF gradients with the $\sigma$DZHF basis ($\vec{F}^\text{HF}_{\sigma \text{DZHF}}$) or analytic gradients with the pcseg-2 basis ($\vec{F}^{E}_{\text{pcseg-2}}$).
    All errors are relative to a reference geometry optimized with analytic gradients and the pcseg-4 basis. 
    Three error metrics are shown: distances between bonded atoms $d$ in \AA{}, angles between bonded atoms $\theta$ in degrees, and dihedral angles $\phi$ in degrees.
    The minimum and maximum absolute errors in all three metrics relative to the reference geometry are provided.
    Linear molecules have no data for $\theta$, and most simple molecules have no data for $\phi$.
    }
    \label{tab:tabl3}
\end{table*}

A comparison of the HF forces with the $\sigma$DZHF basis ($\vec{F}^\text{HF}_{\sigma\text{DZHF}}$) and analytical forces in the $\sigma$DZHF, pcseg-2 and aug-pcseg-2 basis sets ($\vec{F}^{E}_{\text{basis}}$) against reference analytical forces in the aug-pcseg-3 basis set ($\vec{F}^{E}_{\text{aug-pcseg-3}}$) is shown in \figref{dna}.
The distribution of force errors pertaining to the multiple studied configurations is shown with standard box plots, where the box indicates the first to third quantiles, the whiskers denote 90\% confidence interval, and the center line denotes the median error.

We begin by noting that there is little difference between the errors in the pcseg-2 and aug-pcseg-2 analytical forces.
The lack of difference is likely due to the non-equilibrium nature of the DNA MD samples, which had 26 meV additional energy per mode, washing out the importance of diffuse functions in describing the electronic structure.
In contrast to our results, equilibrium calculations on DNA have indicated the importance of diffuse functions particularly in describing the ionic phosphate group.\cite{B600027D, doi:10.1021/ja026759n}
Due to the negligible difference between the pcseg-2 and aug-pcseg-2 forces on our non-equilibrium configurations, we limit the discussion of force errors to the $\sigma$DZHF and pcseg-2 basis sets.

We find excellent agreement for forces computed with the $\sigma$DZHF and pcseg-2 basis sets across all atomic species.
Analytic forces computed with $\sigma$DZHF have comparable error to those computed with pcseg-2, while $\sigma$DZHF HF forces have slightly larger errors than pcseg-2 analytic forces.
To quantify the larger errors present in the HF force calculations, we compute the increase in the median errors shown in \figref{dna} relative to the pcseg-2 analytic forces.
For the H, C, N, O and P atoms, we find the median errors for the $\sigma$DZHF HF forces are larger by a multiplicative factor of 1.5, 1.8, 1.6, 1.1 and 1.9 relative to the pcseg-2 analytic forces---all below a factor of 2.
As the quality of the $\sigma$NZHF basis set is contingent on the quality of the starting basis set, systematic improvements to the HF error can be pursued by beginning with a higher quality starting basis set than the $\sigma$NZ bases used in this work.
Especially, the $\sigma$NZ basis sets have been optimized for atomic CISD energies; starting from a basis set optimized for DFT energies would likely yield smaller errors at lower computational costs for the present purposes of reproducing DFT forces, following the rationale for the polarization consistent basis sets of \citet{Jensen2001_JCP_9113}.
Alternatively, the design of the $\sigma$NZ basis sets could be revisited by adding more functions.

Finally, we note that the error distribution for oxygen is considerably wider than that for the other atoms. 
This can be tentatively explained by the wider range of environments for oxygen in the DNA structures: there is neutral \ce{O^0} in the A, C, G and T bases and a negatively charged \ce{O-} anion in the phosphate group, while the H, C, N and P atoms are only found in their electrically neutral forms in DNA.\cite{watsoncrick1953dna}

\subsection{Geometry optimization with HF forces\label{sec:optimization}}
To demonstrate the usefulness of HF forces, we run geometry optimization on the benchmark set of molecules of \secref{benchmark} using both HF forces and analytic forces.
The starting geometry for all calculations is the non-equilibrium geometry for which forces were calculated in \secref{benchmark}.
For comparison, reference geometries were optimized using analytic forces with the pcseg-4 basis.
We emphasize an important point: geometry optimization using HF forces makes no use of analytic forces, neither in generating the geometry updates, nor in the convergence criteria.

The results of geometry optimization are presented in \tabref{tabl3}.
For each system, we compare two different geometries: an optimized geometry using pcseg-2 analytic forces $\vec{F}^{E}_{\text{pcseg-2}}$, and an optimized geometry using the $\sigma$DZHF basis and HF forces $\vec{F}^\text{HF}_{\sigma \text{DZHF}}$.
We then compute three different metrics comparing the geometry to the reference optimized geometry.

The first metric, $|d - d_{\text{ref}}|$ presents the absolute error in pairwise distances for bonded atoms in a given geometry, $d$ relative to the reference geometry $d_{\text{ref}}$.
To compute this metric we take the given geometry and compute a list of pairwise distances between all pairs of bonded atoms.
An absolute difference between this list of distances and a similar list for the reference geometry is computed, and the minimum and maximum absolute errors in m\AA{} are reported. 
The second metric, $|\theta - \theta_{\text{ref}}|$ follows a similar philosophy,
where we compute a list of angles between bonded atoms for a given geometry and the reference geometry, $\theta$ and $\theta_{\text{ref}}$, and we report again the minimum and maximum absolute difference in degrees.
Finally, we have the third metric $|\phi - \phi_{\text{ref}}|$, where we compute all dihedral angles for a given geometry and the reference, $\phi$ and $\phi_{\text{ref}}$, and report the minimum and absolute difference in degrees.

We begin with geometries optimized with analytic gradients and the pcseg-2 basis.
Pairwise distances agree with the reference within 1 m\AA, with the only exceptional outlier being hydroxysulphane (HSOH) with a 5.756 m\AA{} error for the S--O bond, constituting a small 0.38\% error in the bond length.
For angles, both bonding $\theta$ and dihedrals $\phi$, we find agreement with the reference geometry within a few tenths of a degree, the largest differences being observed in the angles for hydroxysulphane that cap out at 0.490 degrees error in the dihedral angle.

We move next to geometries computed using the $\sigma$DZHF basis and HF forces, which are remarkably accurate.
Perhaps as expected, geometries optimized using the $\sigma$DZHF basis with HF forces yield slightly (2--3 times) larger errors across the board compared to geometries optimized using analytic forces.
Nonetheless, general trends for the optimized geometries using HF forces are promising.
Bond lengths are accurate to a few m\AA{}, and bond and dihedral angles are accurate to a few tenths of a degree.
We see then clearly that one can conduct state-of-the-art geometry optimization with HF forces using the $\sigma$DZHF basis for all molecules considered in this work.

\subsection{Molecular dynamics with HF forces \label{sec:md}}
As a final demonstration of the value of HF forces with the $\sigma$NZHF bases, we conduct a BOMD calculation in the NVE ensemble on a single ethanol molecule using HF forces in the MD integration.
We use the starting configuration for ethanol from the benchmark set of molecules in \secref{benchmark} and impart a randomized initial velocity on each atom in the molecule to provide an average temperature near 300K.
The instantaneous temperature is calculated with the conventional formula 
\begin{equation}
    T(t) = \frac{2}{3(N_\text{atom}-1)k_B} E_\text{kin}(t) 
    \label{eq:temp}
\end{equation}
where $E_\text{kin}$ is the kinetic energy of the molecule, $N_\text{atom}$ is the total number of atoms, $k_B$ the Boltzmann constant, and $t$ is the time.

\begin{figure}
\includegraphics{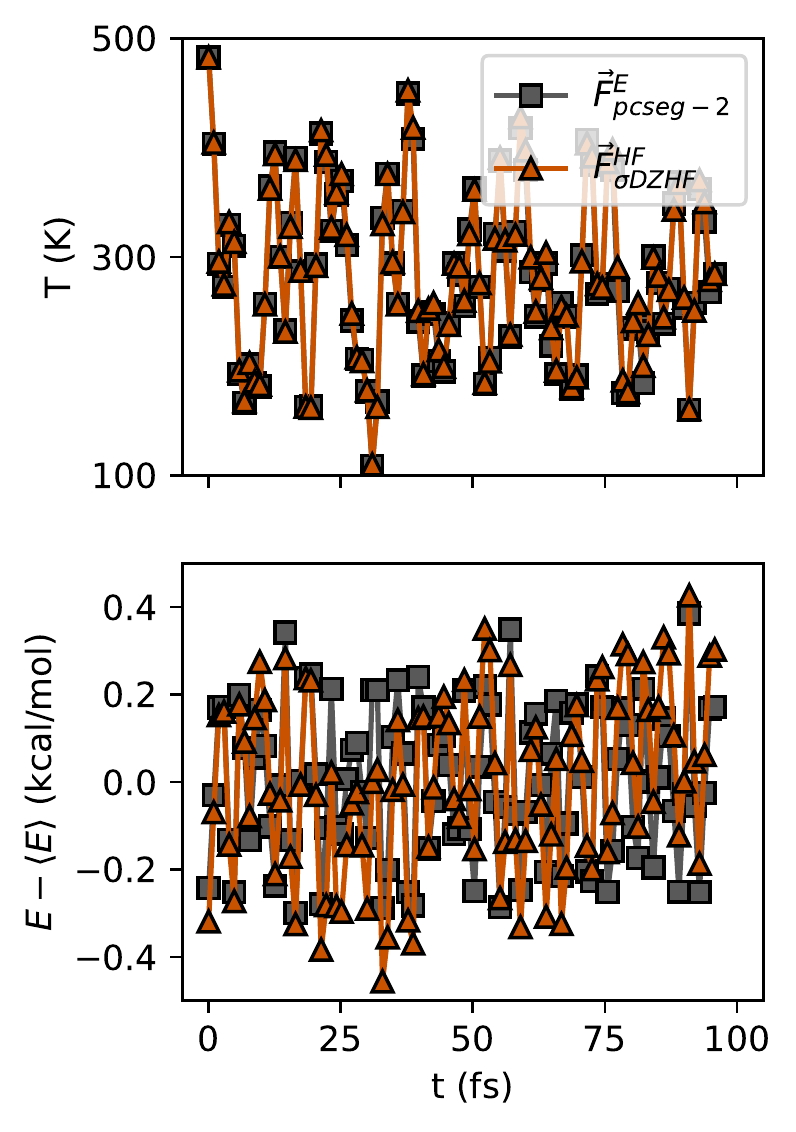}
\caption{Time series data of instantaneous temperature and total energy fluctuations in NVE BOMD simulations of ethanol.
Two results are shown per figure, one for an MD simulation with analytic forces computed using the pcseg-2 basis ($\vec{F}^E_{\text{pcseg-2}}$) and one for an MD simulation with HF forces computed using HF forces and the $\sigma$DZHF basis ($\vec{F}^\text{HF}_{\sigma \text{DZHF}}$).
The time series averaged temperatures $\langle T \rangle$ for the pcseg-2 and $\sigma$DZHF calculations are 280.63 K and 283.76 K, and time series averaged energies $\langle E \rangle$, respectively, $-97200.70$ kcal/mol and $-97207.26$ kcal/mol.}
\label{fig:md_energies}
\end{figure}

\begin{figure*}
\includegraphics[width=\textwidth]{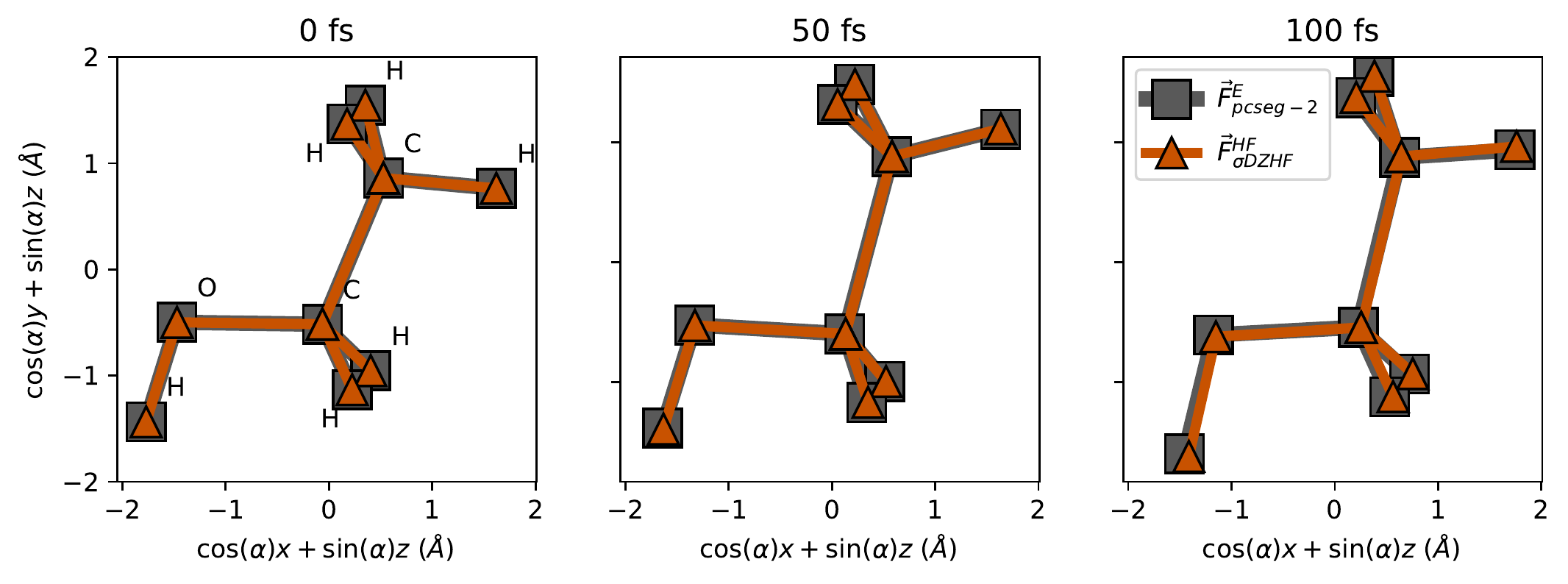}
\caption{Snapshots at 0, 50 and 100 fs comparing between NVE BOMD simulations of ethanol using analytic and HF derivatives.
Transformed coordinates are used to visualize all the atoms in the ethanol molecule in a 2-d plane by rotating the $x$ and $y$ coordinates into the $z$ plane via an angle $\alpha = 5.73 ^\circ \ (0.1$ rad$)$.
A full animation of the MD simulations can be found in the Supplementary Material.
(Multimedia view)}
\label{fig:md_xyz}
\end{figure*}

\Figref{md_energies} shows times series data of the temperature and total energy of the MD calculations of ethanol using HF forces.
As a point of comparison we also include MD simulations computed using analytic forces and the pcseg-2 basis, holding all other details of the MD calculation fixed.
The temperature data shows typical fluctuations in time around a time series average temperature near $\langle T \rangle \sim 300$K as expected.

More interesting is the total energy time series, which we plot as fluctuations around the time series average $\langle E \rangle$.
Conceptually, the fluctuations present in the MD calculation with analytic forces arise from the finite timestep of 1 fs chosen in the Verlet integration.
In addition to the finite timestep error, the MD calculation with HF forces has an additional source of total energy fluctuation arising from the HF forces not strictly satisfying momentum conservation discussed in \secref{diatomic}, potentially leading to spontaneous heating within the simulation.
For both calculations we find fluctuations in the total energy near 0.5 kcal/mol arising from the finite precision of the MD simulation, constituting a relative fluctuation near $10^{-4}$\%, alleviating concerns about erratic behavior in total energy for NVE BOMD calculations with HF forces.

\Figref{md_xyz} shows three snapshots of the ethanol molecule at 0 fs, 50 fs, and 100 fs from our MD calculations.
We plot the configurations using a transformed set of coordinates, since some atoms in the ethanol have identical $x, y$ coordinates and differ only in the $z$ coordinate.
The transformed coordinates are simply rotations in the $x-z$ and $y-z$ planes by an angle $\alpha = 5.73 ^\circ \ (0.1$ rad$)$, which was chosen for visual clarity.
We find excellent agreement between the MD configurations generated using HF forces with the $\sigma$DZHF basis and those generated with analytic forces and the pcseg-2 basis, further solidifying the potency of HF forces with the $\sigma$NZHF basis for accurate MD calculations.
An animated version of the MD simulation with frames at each 1 fs timestep are provided in the Supplementary Material.
 
\section{Summary and Conclusions \label{sec:conclusions}}
We have proposed an algorithm to augment Gaussian basis sets for improved fulfillment of the Hellmann--Feynman (HF) theorem by following the procedure of \citet{rico2007generation}.
We have demonstrated the algorithm with the $\sigma$NZ basis sets of \citet{Ema2022}, resulting in the $\sigma$NZHF basis sets used in the demonstrations of this work.
We computed HF forces and optimized geometries of a large set of molecules with the $\sigma$NZHF basis sets, and found excellent agreement with calculations using analytic forces and state-of-the-art Gaussian basis sets. We also ran a 100 fs BOMD simulation of an ethanol molecule using HF forces computed with the $\sigma$DZHF basis, and found that the configurations and total energies agreed well with a simulation with analytic forces computed with the pcseg-2 basis set.
We have thus demonstrated that the Pulay force can be suppressed in Gaussian basis set calculations, and that forces computed from the HF theorem can be made as accurate as analytical forces from calculations with state-of-the-art pcseg-N Gaussian basis sets.
Our results alleviate long-held concerns regarding the accuracy of HF forces computed with atom-centered basis sets in applications like geometry optimization and MD.\cite{Pulay1977, doi:10.1063/1.446089}

Additionally, by demonstrating that accurate forces can be computed just from an accurate electronic density in line with the Hohenberg--Kohn theorems,\cite{Hohenberg1964_PR_864} our work shines light on an interesting path for first principles ML force calculations.
Instead of focusing on ML models that directly predict the force, one could train an ML density model on first principles densities computed with an appropriate basis set like $\sigma$DZHF or $\sigma$TZHF.
Forces can then be computed directly from the predicted ML density with the HF theorem (\eqref{hft}) and used in practical applications, such as geometry optimization and molecular dynamics.
The primary speedup of this proposed ML pipeline lies in the generation of training data, as computing the electronic density from first principles is much simpler than computing the analytic force. \cite{doi:10.1063/1.1286598, doi:10.1063/1.1562605}
Such an application of ML would enable accurate modeling for large-scale systems that are inaccessible to traditional first-principles techniques.

There are already many ML models which can accurately predict densities.\cite{doi:10.1021/acscentsci.8b00551, doi:10.1021/acs.jcim.1c00227, https://doi.org/10.48550/arxiv.2201.03726}
However, these models typically predict the density in an auxiliary basis set.
As such, the last major hurdle to integrating our accurate basis set to ML HF forces is the construction of optimized \textit{auxiliary} basis sets according to the ordinary basis sets constructed in this work.
With these optimized auxiliary basis sets in hand---which might be generated automatically~\cite{doi:10.1021/acs.jctc.1c00607}---a promising pipeline would emerge for the accurate computation of forces for large-scale systems built on an ML model for the electronic density.
We hope to follow up with work of this nature in the near future.

As a final point, although we used the $\sigma$NZ basis sets of \citet{Ema2022} as the starting point of this work, the algorithms presented herein could also be used with other types of Gaussian basis sets, with the drawback that the resulting HF augmented basis sets will likely be computationally less efficient than the $\sigma$NZHF basis sets employed in the present work, due to the special family-style structure used in the $\sigma$NZ basis sets.

\section*{Supplementary Material}
We provide a detailed description of the basis set augmentation technique, the $\sigma$SZHF, $\sigma$DZHF and $\sigma$TZHF basis sets developed in this work in plain text, and an animation of the ethanol BOMD simulation as supplementary materials. 

\section*{Acknowledgements}
J.A.R., S.P., and A.J.L. were supported by the Harry S. Truman Fellowship, and the Laboratory Directed Research and Development and Academic Alliance Programs of Sandia National Laboratories. Sandia National Laboratories is a multimission laboratory managed and operated by National Technology \& Engineering Solutions of Sandia, LLC, a wholly owned subsidiary of Honeywell International Inc., for the U.S. Department of Energy’s National Nuclear Security Administration under contract DE-NA0003525. We thank the Sandia Academic Alliance for supporting this work. S.L. acknowledges funding from the Academy of Finland through project numbers 350282 and 353749. We thank Sandia National Laboratories and the UNM Center for Advanced Research Computing, supported in part by the National Science Foundation, for providing the high performance computing and large-scale storage resources used in this work. This paper describes objective technical results and analysis. Any subjective views or opinions that might be expressed in the paper do not necessarily represent the views of the U.S. Department of Energy or the United States Government.

\section*{Data Availability Statement}
The $\sigma$NZHF basis sets are made available in the Supplementary Material and online in the Basis Set Exchange \cite{doi:10.1021/acs.jcim.9b00725, doi:10.1021/ci600510j}, and a full animation of the MD simulation is presented in the Supplementary Material as well.
All other data used in this work is available on request from the authors.

\bibliography{main}% Produces the bibliography via BibTeX.
\end{document}